\documentclass[a4paper,10pt]{article}

\usepackage{amsmath}
\usepackage{subfigure}
\usepackage{amssymb}
\usepackage{mathptmx}
\usepackage{graphicx}
\usepackage{multirow}
\RequirePackage{natbib}
\usepackage{eurosym}
\usepackage{rotating}
\usepackage{subfigure}
\usepackage{floatflt}
\usepackage{url}
\usepackage[table]{xcolor}
\newcommand{\ve} {{\varepsilon}}
\newcommand{\R} {{\mathbb R}}
\newcommand{\Tfrac}[2]{\textstyle\frac{#1}{#2}}
\newcommand{\SSs} {\scriptscriptstyle}
\newcommand{\EM} {{\mathbb I}}
\newcommand{\LO} {\mathop{\rm O}\nolimits}
\newcommand{\Lo} {\mathop{\rm o}\nolimits}
\newcommand{\VaR} {\mathop{\rm VaR}\nolimits}
\newcommand{\OpVaR} {\mathop{\rm OpVaR}\nolimits}
\newcommand{\Ew}{\mathop{\rm {{}E{}}}\nolimits}

\date{}

\title{\begin{Huge}Robust Estimation of Operational Risk\end{Huge}
\thanks{
The authors would like to thank Algorithmics~Inc.\ for providing the operational risk data.\newline
Parts of this article are based on the same-named paper presented at the $4^{th}$ European Risk Conference ``Perspectives in risk management: Accounting, Governance and Internal Control'', 14--15 September 2010, Nottingham, UK.\newline
This work is supported by the German Academic Exchange Service (DAAD) for the first author.
}
}

\author{\textbf{Nataliya Horbenko}\\
Fraunhofer ITWM, Department of Financial Mathematics,\\
Fraunhofer-Platz 1, D-67663 Kaiserslautern, Germany, and\\
Dept. of Mathematics, University of Kaiserslautern,\\
\ttfamily Nataliya.Horbenko@ITWM.Fraunhofer.de,\\
\ttfamily +49 \ 631/31600-4333
\and
\textbf{Peter Ruckdeschel}\\
 Fraunhofer ITWM, Department of Financial Mathematics,\\
Fraunhofer-Platz 1, D-67663 Kaiserslautern, Germany,\\
\ttfamily Peter.Ruckdeschel@ITWM.Fraunhofer.de,\\
\ttfamily +49 \ 631/31600-4699
\and
\textbf{Taehan Bae}\\
 Algorithmics~Inc., 185 Spadina Avenue, Toronto,\\
Ontario M5T 2C6, Canada,\\
\ttfamily Taehan.Bae@Algorithmics.com\\
\ttfamily +1 \ 416/217-4436}

\begin{document}

\maketitle

\newpage

\textit{
According to the Loss Distribution Approach, the operational risk of a bank
is determined as the $99.9\%$ quantile of the respective loss distribution, covering unexpected
severe events. The $99.9\%$ quantile can be considered a tail event.
As supported by the Pickands-Balkema-de Haan Theorem, tail events exceeding some high threshold
are usually modeled by a Generalized Pareto Distribution (GPD).
Estimation of GPD tail quantiles is not a trivial task, in particular if one takes
into account the heavy tails of this distribution, the possibility
of singular outliers, and, moreover, the fact that data is usually
pooled among several sources.\newline
In such situations, robust methods may provide stable estimates when classical methods already fail.
\newline
In this paper, optimally-robust procedures MBRE, OMSE, RMXE are introduced to the application domain of operational risk.
We apply these procedures to parameter estimation of a GPD at data from Algorithmics~Inc.
To better understand these results, we provide supportive diagnostic plots adjusted for this context:
influence plots, outlyingness plots, and QQ plots with robust confidence bands.
}

\vspace*{3ex}


\section*{Introduction}

Operational risk according to \cite{Basel:06} covers risks of loss resulting
from inadequate or failed internal processes, people or systems, or from external events.
Neither is this risk new, nor is the need to measure it. Still, it remains a challenging
issue, in particular as far as very large operational losses are concerned, compare \cite{deFon:06}.
This is reflected by the sizeable amount of media coverage we have seen in the last few years:
rogue tradings, e.g., at Daiwa Bank (1984-95), Sumitomo Corp. (1986-96),
Barings (1995), and Soci\'et\'e G\'en\'erale (2006-2008), losses caused by
the 9/11 terrorist attacks (2001), by the B.~L. Madoff fraud (1980s-2008),
by hurricane Katrina (1995), and by the recent earthquake in Japan (2011).
Because of its impact, operational risk has been integrated into the Basel~II
framework of regulatory requirements.
The focus of the present paper lies in (robust) quantification of the respective
regulatory capital.

One of the most challenging problems in this context is data---both as
to quantity and as to quality.
We notice that, fortunately for our economies, very large operational losses are
observed rarely. Still, they have a tremendous effect.
As a consequence, usually only some few observations will have an overwhelming
impact on the computed regulatory capital. In addition, in a realistic modeling,
taking into account possible model deviations, one cannot tell (without error)
whether these events are singular outliers or reproducible and, hence, contribute
valuable evidence for future losses. This question of relevance for future losses
gets even more severe in the common and Basel-II-recommended practice
of data pooling used to overcome the lack of historical (very large) loss data.

Let us illustrate this: in quantifying risk, usually the tail
behavior of the underlying distribution as expressed by tail quantiles
(VaR) or truncated moments (CVaR) is crucial.
Estimating these population quantiles by their empirical counterparts
apparently is drastically prone to outliers: for the $99.9\%$
quantile as typically used in operational risk,
for 5000 observations, five irreproducible, extra-ordinarily large observations
suffice to render this procedure completely meaningless.
Passing to parametric models from extreme value theory per se is no remedy:
maximum likelihood estimators (MLEs), optimal in this context, as a rule
still attribute unbounded influence to some exposed observations,
e.g., in our example, five outliers will still suffice to invalidate
our conclusions.

This is where robust statistics steps in.
It aims at designing procedures which remain stable under minor model
deviations; these deviations can stand for a minority of unpredictable
outliers for which we cannot anticipate any model distribution.
In our little illustration, robust statistics provides procedures bounding
the influence of single observations.

\medskip

While \cite{C:R:06} have introduced general robust concepts
to the domain of operational risk, the contribution of this paper
is the application of \textit{optimally robust} procedures
to the quantification of operational risk,
more precisely to data from the \texttt{Algo OpData} database of
Algorithmics~Inc.
To this end, we focus on the part of operational risk caused by
very large losses; i.e., on the tail distribution of
the severity of operational losses, which leads us canonically to
(optimally-robust) parametric estimation in generalized Pareto
distributions.

To this end, we present a comprehensive, self-contained survey of the
shrinking neighborhood setup, in which the respective
optimally-robust estimators are derived.
We do not repeat the respective derivation here, nor do we conduct a
simulation study, comparing them to competitor estimators
which could support our findings also for finite sample sizes.
Instead, we refer the reader to \cite{R:H:10} which contains all this.

To judge the quality of our estimators when applied at real data
sets (where fulfillment of the actual model assumptions is not clear),
we contribute the translation of some diagnostic plots from
robust statistics to this application domain which help us to
understand and quantify the effect of our robustifications.

\medskip

The rest of the paper is organized as follows: the setup
is presented in detail in Section~\ref{Setup}, starting
 with the regulatory framework, describing the data situation,
defining the mathematical setup in two parts as to the Loss
Distribution Approach and the generalized Pareto distribution
to model the tail of the severity distribution.
Section~\ref{Robustness} continues with robustness: after an introduction of the
central concepts of robust statistics we give a short summary
of the literature on robustness approaches relevant for operational risk.
Its longest subsection, Subsection~\ref{Sec:Estimators}, contains the
announced self-contained summary of the shrinking neighborhood approach
of robust statistics, in which we have obtained the optimally-robust estimators
OMSE, MBRE, and RMXE used in the sequel. At the end of this section,
we provide some implementation details.
In Section~\ref{Sec:DataSet}, the data set from Algorithmics~Inc.\ is discussed,
together with the evaluation of the considered estimators at this
data.
Section~\ref{Sec:Diagnostics} finally provides the diagnostic plots, which
are again produced and explained for data from the \texttt{Algo OpData} database.
A conclusion section at the end summarizes our main findings.

\section{Setup}\label{Setup}
\subsection{Regulatory Framework}\label{RegF}

The most important international set of regulatory rules
for financial institutions is given by
the Basel~II framework for the International Convergence  of
Capital Measurement and Capital Standards (\cite{Basel:06}),
which in particular covers operational risk.

The question to which Basel~II applies is currently an
important political one, but not the topic of this article.
We only note that Basel~II is binding for all financial services
institutions in the  European Union since 2007,
but so far only covers the largest or most internationally active banks in the USA,
a situation to be changed only in the upcoming Basel~III framework targeted for implementation in 2013.
The results of a survey of the Basel Committee on Banking Supervision (\cite{Basel:10a})
indicate that 112 countries have implemented or are currently planning to implement Basel~II.

According to the Basel~II framework, every bank has to estimate its
operational risk and hold the appropriate regulatory capital to ensure
its solvency and economic stability in case of foreseeable operational
losses. While Basel~II rules  mainly address large, internationally active banks,
their basic concepts should be applicable to banks of varying organizational
and product line complexity.

Basel~II further recommends certain approaches for measuring the
operational risk: the \textit{Basic Indicator Approach}, the
\textit{Standard Approach}, and the \textit{Advanced Measurement Approaches (AMAs)}.
The most sophisticated approaches are gathered in group AMA, which is advised for
large international banks, but also subject to supervisory approval (\S 655, \cite{Basel:06}),
for which a bank must meet certain qualitative and quantitative standards.
The focus of this paper lies on the Loss Distribution Approach (LDA),
which is a particular AMA to be discussed in Subsection~\ref{Sec:LDA}.

\subsection{Data Situation}\label{datasit}
LDA suggests measuring the operational risk based on historical
data using information about the frequency and severity of earlier losses.
To this end, according to Basel~II, past operational losses of a bank
should be documented in internal databases.

\begin{figure}
\begin{center}
\fbox{\includegraphics[scale=0.33,angle=270]{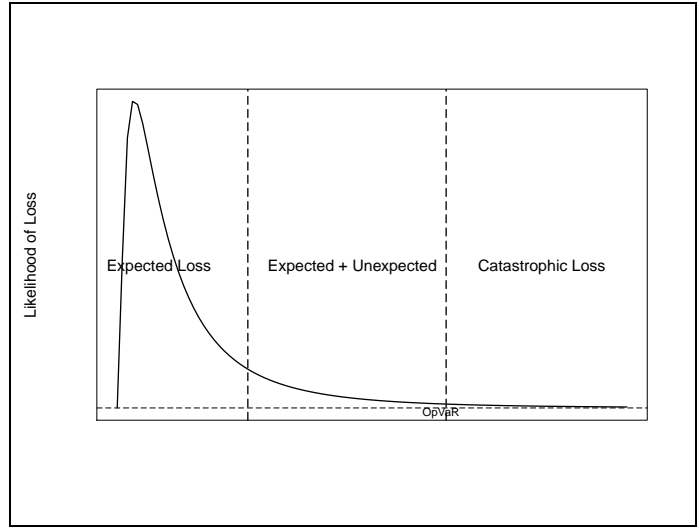}}
\label{Fig:Losses}
\end{center}
\caption{\footnotesize Loss Categories}
\end{figure}

These losses can roughly be divided into three types: expected
(occasional and moderate), unexpected (rare, but large), and catastrophic
(very rare, extreme) losses (see Figure~\ref{Fig:Losses}), where,
according to  \cite[\S 669 (b)]{Basel:06}, the regulatory capital is obtained
as the sum of expected and unexpected losses.
As mentioned in the introduction, unexpected losses are rare events,
so the data situation is most difficult for this segment.

As internal time series of this type are usually short and sparse, external data
from losses of other banks documented in publicly available data pools
(such as Algorithmics \texttt{Algo OpData}, SAS OpRisk Global Data) or
databases of consortia of banks (e.g., ORX), as well as scenario-based data and
internal control factors, should be included into estimation of
the regulatory capital (\cite{Basel:10}).
Inclusion of external data introduces new statistical challenges:

First of all, there is the question of size and comparability; \`a priori
it is far from clear whether losses of one bank could occur at all at another bank,
and if so, at which scale. Scaling of external data to a particular bank is a topic
in its own right and has been dealt with in detail in \cite{C:L:08}, and \cite{C:J:Y:11};
we do not go into this in this article. We only note that even after a proper and
robust scaling step, the robustness issue is not yet removed.

In addition, we face a censoring problem, especially for external data,
since data usually is only reported beyond a certain threshold.
For internal reporting, this threshold is usually
set relatively low, e.g., at {10,000}~EUR according to a Basel~II suggestion.
For losses reported to the outside world, it varies from EUR~{20,000} (ORX) to
1 million USD (\texttt{Algo OpData}). So in particular external loss samples
will be biased, containing  disproportionally high numbers of very large losses
(\cite{deFon:06}).

\subsection{Mathematical Setup I: Loss Distribution Approach}\label{Sec:LDA}

As indicated, this paper focuses on the Loss Distribution Approach (LDA).

In this approach, banks estimate the operational risk separately for each of
some eight business lines and seven event types, giving a partition
into a matrix, see Table~1. The cells of this matrix are not
stochastically independent, which is relevant for the aggregation
of these cells to a total operational risk exposure; to this end,
the cell  dependence structure is usually captured by copula techniques.
In this paper, however, we skip this aggregational step and
rather confine ourselves to cell-individual results.

 As in the Collective Model in actuarial context, in LDA,
 severity and frequency of the operational losses
 are modeled separately, and total loss is determined as the
 compound distribution; i.e., given the distribution of
 the frequencies $N$ within time period $t$ and distribution $F$ of loss
 severities $X_i$, the aggregate or cumulative loss $L$ over $t$ is
 calculated as

$$L_t = \sum_{i=1}^{N(t)} X_i\;.$$

In this context, annual frequency of losses are usually modeled by
Poisson or the negative binomial distribution (\cite{Mosc:04}).
Assuming independent and identically distributed severities $X_i\sim F$,
$L_t$ is said to follow a compound process with distribution
${\cal L}(L_t)$ given by its cumulative distribution function (cdf)

 $$\mathbb{P}(L_t \leq x) = \sum_{k=0}^{\infty}\mathbb{P}(N(t) = k)
    \cdot F^{\ast k}(x), \quad k \in \mathbb{N}$$

 where $F^{\ast k}(x)$ is the cdf of the $k$-fold convolution of $F$.
 The regulatory capital is determined applying a risk measure, e.g.\
 Value-at-Risk ($\VaR$), to  ${\cal L}(L_t)$. The operational
 Value-at-Risk ($\OpVaR_\alpha$) is the corresponding $\alpha$ quantile
 of ${\cal L}(L_t)$  as required by Basel~II.
 Computation of this compound distribution ${\cal L}(L_t)$  can be
 tackled by simulations, approximations, and other techniques---we do not detail this here.

Generally, loss data can be fitted to a variety of severity distributions:
medium-tailed Exponential, Lognormal, Gamma, Gumbel; heavy-tailed Pareto,
GPD, Burr, Loggamma,  Weibull with shape $\xi<1$.
Basel~II requirements stress that a bank must be able to
demonstrate that its approach captures `tail' events (\S 667, \cite{Basel:06}).
As discussed by \cite{deFon:07}, there is evidence for individual
operational losses of banks which are heavy-tailed with existing first
but infinite second moments; moreover, for pooled data even the first moments
may not exist (\cite{Mosc:04}).

If the underlying severity distribution $F$ is subexponential\footnote{
A distribution $F$ is subexponential iff
$\overline{F^{\ast n}}(x) \approx n \overline{F}(x), \quad x \rightarrow \infty$,
where $\overline{F^{\ast n}}$ is the survival function of an $n$-fold convolution of $F$.}, \cite{B:K:05}
show the validity of the following first-order approximation for a
high quantile of the compound distribution,
the so-called single-loss approximation:

\begin{equation}\label{Eq:OpVaR}
\OpVaR_\alpha \approx F^{-1}\left(1-\frac{1-\alpha}{\lambda t} \right),
   \quad \alpha \rightarrow 1\;,
\end{equation}
where $\lambda = \mathbb{E}(N(t)) / t$ is the expected frequency per unit of time $t$, and is the rate or intensity of the Poisson (point) process of loss events; $F^{-1}$ is a corresponding quantile function.
For Poisson distributed $N(t)$, we note that the MLE for $\lambda$ is just
the average number of losses over time period $t$.
As in practice all commonly used heavy-tailed distributions belong to
the subexponential class, it is usually enough to estimate the quantile
of severity distribution of losses only.

\subsection{Mathematical Setup II: Generalized Pareto Distribution}\label{Sec:GPDMod}

As we are interested in the tails of the severity distribution, extreme
value theory (EVT) applies, providing models for rare and extreme events,
see \cite{Chav:06,Emb:03,Nesl:06}.

One of the most prominent results of EVT, the Pickands-Balkema-de Haan
Theorem  (see \cite{B:H:74,Pick:75}), states that if the distribution of the standardized maxima
of $X$ tends to an extreme value distribution, the peaks over a high
threshold $u$ are asymptotically distributed as a generalized Pareto
distribution (GPD) $G_{u,\xi,\beta}$:

$$\mathbb{P}(X-u \leq x| X>u) \approx G_{u,\xi,\beta}(x), \ x > u \;.$$

This gives rise to the so-called \textit{Peaks Over Threshold} method
(POT) and motivates the use of the GPD for modeling the tail of the
severity distribution, provided threshold $u$ is chosen appropriately.

Limitations of this motivation are given by its asymptotic
nature and its applying for extremal order statistics only.
Hence, to obtain thresholds for which this motivation applies, one has
 to find a suitable trade-off between the lack of data beyond
 this high threshold and a large deviation from the asymptotic
 distribution.

\paragraph{Parameters of the GPD}
The GPD is specified through its cdf:

$$G_{u,\xi,\beta}(x) = 1- \left(1+\xi \frac{x-u}{\beta}
                          \right)^{-1/\xi}, \; x>u .$$

In the GPD, the shape parameter $\xi$ controls the form
of the distribution; more specifically, only values $\xi > 0$ are of
interest in our context, as otherwise the support of this distribution
will be bounded. $\beta>0$ is a scale parameter and $u$ is a location parameter,
which acts as threshold, usually unknown. Estimation of $u$ is a difficult task,
as standard methods from smooth parametric statistics do not apply.
Several approaches, using criteria such as minimum mean prediction
error (or robust variants) or minimum squared error, are around though, compare
\cite{B:V:T:96,B:D:G:M:99,Du:98,D:V-F:06,V:B:C:H:07}.

Note that the underlying distribution of $X$ can be approximated as:

$$\hat{\overline{F}} (x) = \frac{N_u}{n}\left(1-G_{u,\hat{\xi},\hat{\beta}}(x)
   \right)\;,$$

where $n$ is a total sample size, $N_u$ is the number of exceedances over
the threshold $u$, $\overline{F}(x) = 1- F(x)$ is the survival function.
Applying (\ref{Eq:OpVaR}), operational $\alpha$-Value-at-Risk of a
compound loss then is merely a corresponding $\alpha'$-quantile
of $\hat{F}$ and equal to
$$
\OpVaR_{\alpha} = u +\frac{\hat{\beta}}{\hat{\xi}} \left({\alpha'}^{-\hat{\xi}}-1
\right), \ \alpha' = \frac{n}{N_u} \cdot \frac{1-\alpha}{\lambda t} \;.
$$

Estimation of $\theta=(\beta, \xi)$ for given threshold $u$ in GPD models has
widely been studied. A detailed analysis of existing and new methods
for the estimation of GPD---both classical and robust---can be found in \cite{R:H:10}.
This also is the reference model for the remainder of this paper, i.e., for some given $u\in\R$
\begin{equation} \label{GPDmodel}
{\cal P}=\{G_{u,\xi,\beta}\;|\; \beta>0, \xi >0 \} .
\end{equation}


\section{Robust Statistics}\label{Robustness}

Robustness is a stability notion. In robust statistics,
it denotes stability with respect to deviations from the distributional
assumptions, most prominently caused by outliers.
There is a vast body of literature on this topic, starting with \cite{Hu:64},
and with excellent monographs given by, e.g., \cite{Hu:81}, \cite{Ha:Ro..:86},
\cite{Ried:94}, \cite{M:M:Y:06}.

In this section, we compile the necessary concepts and results from
robust statistics needed to obtain the optimally-robust estimators used in this article.
Most of this section holds for general (smooth) parametric
models. The respective terms for model~\eqref{GPDmodel},
though, are spelt out in Section~\ref{Sec:Estimators}.

\subsection{Robustness Concepts}
\label{Sec:Robustness}

Mathematics has long been concerned with stability and offers a set of very
fruitful concepts to operationalize it: continuity, differentiability,
closeness to singularities.

To make these available in our context, it helps to consider an estimator
as a function of the underlying distribution. More precisely, we will consider
functionals $T$ mapping (a subset of all) distributions (at least all model
distributions $F_\theta$) to the parameter set $\Theta$. If we plug in the
true model distribution $F_\theta$ it is natural to require
that $T(F_\theta)=\theta$, i.e., \textit{Fisher consistency}.
In this setting, the estimator is interpreted
as $T$ applied to the empirical distribution $\hat F_n$.

As for growing sample size, in the classical setup, the empirical
distribution will converge to the theoretical one according to
the Glivenko-Cantelli, respectively Donsker Theorems \cite[e.g.\ Thm's.~19.1, 19.3]{VdW:98},
we would expect that a ``good'' functional will respect this convergence
in the sense that also $T(\hat F_n)\to T(F_\theta)=\theta$ suitably.

In particular, weak convergence, as stated in Donsker's Theorem, respectively
closeness in weak topology, help to formulate interesting neighborhoods
of distributions able to capture deviations of distributions or
outlier phenomena.

One of the most practicable types of such neighborhoods is the
\textit{Gross Error Model}: for a given, central distribution $F$ and
a radius $\ve\in[0,1)$, we consider the set (or ball) of all distributions $Q$
obtained as
\begin{equation} \label{nbdef}
{\cal U}=\{Q\,\mid\, Q=(1-\ve)F+\ve H\,\}
\end{equation}
where $H$ is an unknown, uncontrollable, unpredictable outlier generating distribution.

Continuity, more precisely equi-continuity, of a functional on such
neighborhoods (uniformly in growing sample size) is then called
\textit{qualitative robustness} \cite[Sec.~2.2~Def.~3]{Ha:Ro..:86}.

In robust statistics one distinguishes between global and local robustness
of an estimator. Local robustness asks how small deviations, in extreme case
a single observation, influence the value of the estimator.
This is captured by the influence function IF---a functional
derivative\footnote{Strictly speaking in mathematical
terms, this is the G\^ateaux derivative of $T$ into  the direction of the
tangent $\delta_x-F$. To derive certain properties from this differentiability,
in particular asymptotic normality, this notion is in fact too weak, and one has
to apply stronger notions like Hadamard or Fr\'echet differentiability;
for details, see \cite{Fe:83} or \cite[Ch.~1]{Ried:94}.} of the estimator
defined as \cite[Sec.~2.1~Def.~1]{Ha:Ro..:86}:
\begin{equation}
\psi(x):= \mathop{\rm IF}(x;T,F) =
                \lim\nolimits_{\epsilon \rightarrow 0}
                \Big(T((1-\epsilon) F + \epsilon \delta_x)-T(F)\Big)/\epsilon,
\end{equation}
provided the limit exists and where $\delta_x$ denotes the Dirac measure in $x$.
This influence function exactly gives us the infinitesimal influence
of a single observation on the estimator. Under additional assumptions,
one can read off the asymptotic variance of the estimator in the ideal model
as the second moment of $\psi$. Infinitesimally, i.e., for $\ve\to0$,
the maximal bias on ${\cal U}$  is just $\sup|\psi|$, where $|\,\,\cdot\,\,|$ denotes
Euclidean norm. $\sup|\psi|$ is then also called \textit{gross error sensitivity} (GES),
\cite[(2.1.13)]{Ha:Ro..:86}.
An estimator is locally robust iff its GES is finite.

Global robustness of the estimator describes the behavior of the estimator
under massive distortions. It may be quantified by the breakdown point of the estimator---the
maximal radius $\ve$ the estimator can cope with without producing an arbitrary large bias;
it comes with a functional and a finite sample notion, see
\cite[Sec.~2.2~Def.'s~1,2]{Ha:Ro..:86} for formal definitions.
Mathematically this is, hence, nothing but the closest singularity of the max-bias curve.

Robust estimators are constructed to be both globally and locally robust.
This stability comes at the cost of some efficiency in the ideal model:
compared to classically optimal estimator, i.e., the MLE in most cases,
robust estimators are less efficient as quantified by the \textit{asymptotic
relative efficiency} (ARE), i.e., the ratio of the respective two (traces of the)
asymptotic (co)variances, which is strictly smaller than $1$ as a rule,
while a (maximal) value of $1$ would indicate that we attain the same
accuracy as the (classically) optimal estimators. Such an estimator would be
called \textit{efficient}.

\subsection{Robust Methods for Operational Risk}
As detailed in Subsection~\ref{datasit}, data is an important issue in
estimation of operational risk. This issue, by arguments as in \cite{C:R:06},
can be approached by robust statistics. In particular this helps controlling
the bias induced by outliers, censoring, and data heterogeneity, which
can result in systematic over- or underestimation of operational risk.
From a regulatory perspective, underestimation is to be
avoided, while overestimation would not be equally harmful. A risk manager,
on the other hand, also has to take into account  opportunity costs when not
investing available capital, so for him overestimation is also an issue.

A common misunderstanding when applying robust estimation to extremes
is that the extremes themselves are outliers. This need not be the case;
in fact, outliers are observations which are not following the general
pattern of data, which is not necessarily connected to size.
From \cite{D:E:09}, we retain three main messages concerning application
of robust methods to extremes: 1)  ``Robust methods do not downweigh extreme
observations if they conform to the majority of data.'' 2) ``Robust methods
can guarantee a stable efficiency, MSE, and a bounded bias over a whole
neighborhood of the assumed distribution.'' 3) ``Robust methods can identify
influential points in real data''.

Applications of robust statistics to extreme value distributions
can be found in, e.g., \cite{F:S:94,Du:98,D:F:98,P:W:01,D:M:02,J:S:04,B:K:09}.

\subsection{Optimally Robust Estimation---Applied to GPD}
\label{Sec:Estimators}
To operationalize robust estimation, quality criteria are needed,
which summarize the behavior of an estimator on a whole neighborhood,
as in \eqref{nbdef}. In this context two canonical criteria for
parameter estimators have emerged: maximal MSE (maxMSE) on some neighborhood
${\cal U}$ around the ideal model and maximal bias (maxBias) on the
respective neighborhood.

\paragraph{Robust Optimality Problems} This gives the following optimization
problems:
$$\mbox{\hspace{-1.3cm}(Opt1)}\qquad \mbox{minimize maxMSE on
${\cal U}$,\hspace{3cm}}\qquad\qquad
\mbox{\hfill(Opt2)} \qquad\mbox{minimize maxBias on ${\cal U}$}
$$

The respective optimal estimators are called OMSE (\textit{O}ptimal
\textit{MSE} estimator) and MBRE (\textit{M}ost \textit{B}ias
\textit{R}obust \textit{E}stimator), respectively.
A variant (Opt1') of (Opt1) separates MSE into bias and variance and requires
$$\mbox{\hspace{0.6cm}(Opt1')}\qquad \!\!\mbox{minimize the variance in the
ideal model subject to a uniform bias bound $b$ on ${\cal U}$\hspace{5cm}}
$$
%
giving OBRE (\textit{O}ptimally \textit{B}ias  \textit{R}obust
\textit{E}stimator)\footnote{The terms OBRE and MBRE are taken from
\cite{Ha:Ro..:86}, while the notion OMSE is coined in \cite{R:H:10}.}, as
discussed in GPD context, e.g., in \cite{D:F:98}.\smallskip\\

{\small \textbf{Remark} Radius $\ve$ and bias bound $b$ can be seen as
tuning parameters determining the degree of robustness. The larger
$\ve$ (smaller $b$) the more robust is the respective optimal procedure.
The most frequently used tuning criterion though is the \textit{Anscombe criterion}
choosing $b$ such that a prescribed ARE, typically $95\%$,
is achieved in the ideal model. This criterion does not
properly reflect the difficulty of the respective robustness problem, however.
Instead, we propose a different criterion yielding estimator RMXE below.
In particular, in the GPD model, for $\xi=0.7$, with the Anscombe criterion,
we may drop down to $14\%$ relative efficiency for sufficiently large radius
when compared to the OMSE, knowing this radius, whereas RMXE (also without knowing the
radius) never drops below $68\%$ in the same criterion.
}

\paragraph{Shrinking neighborhoods} For solving these problems, we note
that, as a rule, bias and variance scale differently
on neighborhoods of size $\ve$ for growing sample size $n$: while variance
usually is $\LO(1/n)$, maximal bias is $\LO(\ve)$ (for robust estimators).
So for growing $n$, with fixed neighborhood size $\ve$, bias will be
dominant eventually in $n$, leading only to problems of type (Opt2).
To balance bias and variance, the shrinking neighborhood approach (see
\cite{Ried:94}, \cite{R:06}, \cite{K:R:R:09}) sets $\ve=\ve_n=r/\sqrt{n}$
for some initial radius $r\in [0,\infty)$.

While in Subsection~\ref{Sec:Robustness}, we have started with
a given procedure and then determined its influence function,
in the shrinking neighborhood approach, optimality is assessed by
determining optimal influence functions and, in a second step then
estimators are constructed which have this optimal influence
function (``uniformly on the shrinking neighborhood'').

One has to admit that the justification of this approach is merely
asymptotic, i.e., for large sample size. Whereas general statements
for finite samples properties are out of reach,
for given estimators these properties can be assessed through simulations:
in the simulation study carried out for the GPD case
in \cite{R:H:10}, the respective asymptotically optimal estimators
remained optimal (among the considered alternatives) down to sample size $n=40$.

\paragraph{ALEs} The key concept behind this is
\textit{asymptotically linear estimators} (ALEs).
In the simplest setting, we start with a smooth ($L_2$-differentiable)
parametric model  ${\cal P}=\{P_\theta,\;\;\theta\in\Theta\}$ for
independent, identically  distributed observations $X_i\sim P_\theta$ with
open parameter domain $\Theta\subset\R^k$, with scores\footnote{Usually
$\Lambda_\theta$ is the logarithmic derivative of the density w.r.t.\
the parameter, i.e., $\Lambda_\theta(x)=\partial/\partial \theta
\log p_\theta(x)$.} $\Lambda_\theta$ and finite Fisher information
${\cal I}_\theta=\Ew_\theta \Lambda_\theta\Lambda_\theta^\tau$.
In this setting, an influence function is \textit{any}
function $\psi_\theta\in L_2(P_\theta)$ with $\Ew_\theta \psi_\theta=0$ and
$\Ew_\theta \psi_\theta \Lambda_\theta^\tau= \EM_k$ where $\EM_k$ is the
$k$-dimensional unit matrix. The set of all such influence functions is
denoted by $\Psi_2(\theta)$.
Then a sequence of estimators $S_n=S_n(x_1,\ldots,x_n)$
is an ALE if
\begin{equation} \label{ALE}
S_n=\theta+\frac{1}{n}\sum_{i=1}^n \psi_\theta(X_i) +
   \Lo_{P_\theta^n}(n^{-1/2})
\end{equation}
for some influence function $\psi_\theta\in \Psi_2(\theta)$ (which is uniquely
specified by \eqref{ALE}). In the sequel we fix the true $\theta\in\Theta$
and suppress it from notation where unambigous.
Note that the set of ALEs covers a huge variety of estimators,
starting from MLEs, M-estimators, Z-estimators, L-estimators, R-estimators,
quantiles, and many more;
in fact, to derive asymptotic normality of an estimator, most frequently a
representation like \eqref{ALE} is shown as an intermediate result.
In particular, the MLE usually has influence function
$\psi^{\rm\SSs MLE}={\cal I}^{-1} \Lambda$.

\paragraph{The GPD case:} \label{GPDParag}
Model~\eqref{GPDmodel} is smooth, i.e.,\ $L_2$-differentiable,
as the density $f_\theta$ is differentiable
in $\theta$ and the corresponding Fisher information is finite and
continuous in $\theta$ \cite[Satz~1.194]{Wit:85},
with $L_2$-derivative
\begin{equation} \label{LBdef}
\Lambda_\theta(z) = \left( \Tfrac{1}{\xi^2}\log(1+\xi z) -
\Tfrac{\xi+1}{\xi}\Tfrac{z}{1+\xi z};-\Tfrac{1}{\beta}+
\Tfrac{\xi+1}{\beta}\Tfrac{z}{1+\xi z}\right)^\tau, \quad
z = \Tfrac{x-u}{\beta}
\end{equation}
and Fisher information ${\mathcal{I}}_\theta$ as
\begin{equation}\label{FIdef}
\mathcal{I}_\theta = \frac{1}{(2\xi+1)(\xi+1)}
\left(\begin{array}{cc}
 2, & \beta^{-1} \\
\beta^{-1}, & \beta^{-2} (\xi+1)
\end{array}\right)
\end{equation}
 As ${\cal I}_\theta$ is positive definite for $\xi>0$, $\beta>0$,
 the model is (locally) identifiable.
 In particular, both coordinates of $\psi^{\rm\SSs MLE}_\theta$ are unbounded,
 which implies that the MLE is \textit{not locally robust},
 as it has infinite GES.

\paragraph{Optimal solutions}
ALEs are of particular interest as many of their asymptotic properties,
can be obtained, even uniformly on neighborhoods, solely based on their
influence functions. For instance, Problem~(Opt1') becomes
\begin{equation}
\mbox{\rm minimize } \Ew |\psi|^2 \qquad \mbox{\rm subject to}\quad \sup|\psi| \leq b, \quad
 \psi \in\Psi_2
\end{equation}

In particular, the influence functions corresponding to OMSE and MBRE,
$\overline{\psi}$ and $\tilde{\psi}$, respectively, are determined in
\cite[Thm.'s 5.5.7 and 5.5.1]{Ried:94} as solutions to the implicit equations
\begin{equation} \label{OMSEdef}
\overline{\psi} = Y \min \left\{1,{b}/{|Y|}\right\}, \quad Y = A \Lambda - a,
\quad r^2 b = \Ew(Y-b)_+ \;,
\end{equation}
in case~(Opt1),  and, similarly, for case~(Opt2) by
\begin{equation} \label{MBREdef}
\tilde{\psi} = b{Y}/{|Y|}, \quad Y = A \Lambda - a, \quad b =
\max_{a,A}\{\mathop{\rm tr A}/{\Ew|Y|}\}\;.
\end{equation}
where $\mathop{\rm tr}(A)$ is the trace of $A$,
$(\,\cdot\,)_+=\max(\,\cdot\,,0)$, and  $A \in\R^{k\times k}$,
$a\in\R^k$, $b>0$ are Lagrange multipliers
ensuring that  $\psi \in\Psi_2$.
\begin{small}
\textbf{Remark} (i) Both $\overline{\psi}$
and $\tilde{\psi}$ are built up from an affine transformation $Y$
of the scores $\Lambda$. The term $b {Y}/{|Y|}$ retains the direction
of $Y$ but clips it to length $b$. For $\overline{\psi}$ this clipping
is done whenever the length of $Y$
is larger than $b$, whereas in $\tilde{\psi}$ one always clips.

(ii) The solution to (Opt1') coincides with the one for
(Opt1), except that instead of the utmost right equation in \eqref{OMSEdef}
to determine $b$, bias bound $b$ is already fixed in advance in (Opt1').

(iii) Both $\overline{\psi}$ and $\tilde{\psi}$
only can become $0$ if $Y=0$, which means that contrary to practitioners'
rules, in the optimally-robust influence functions, observations are
not thrown away when they are ``large'' or when their influence
measured by $|Y|$ is large. At most, their influence gets clipped.

(iv) Insisting on $\psi\in\Psi_2$ also ensures (asymptotic) unbiasedness in the
ideal model, which is not true per se if, in a model of asymmetric distributions
as the GPD, we simply skip the largest observations. As a rule
such estimators obtained from skipping large order statistics will need a
bias correction.
\end{small}

\paragraph{One-step construction} Having determined the optimally-robust
influence functions, we still have to solve the
already mentioned construction problem, i.e., find an estimator achieving
these prescribed influence functions $\psi=\overline{\psi},\tilde{\psi}$.
Several techniques are available, see \cite[Ch.~6]{Ried:94}; for simplicity,
we apply the one-step construction: for some suitably robust and consistent
starting estimator $\theta_0$, such an estimator is defined as
$$S_n = \theta_0 + \frac{1}{n} \sum_{i = 1}^{n} \psi_{\theta_0}(X_i)$$
Then $S_n$ is an ALE with influence function $\psi$. As a key feature, at least
as long as\footnote{For example, in case of scale parameter $\beta$ in the GPD,
restricted to $(0,\infty)$, this can be achieved by a logarithmic transformation
of the parameter space.} $\Theta=\R^k$, the breakdown point properties of the
starting estimator $\theta_0$ are inherited unchanged to $S_n$.

\paragraph{The starting estimator}
 $\theta_0$ in this construction is required to be sufficiently smooth and
 to be of accuracy $\LO_{P_\theta^n}(1/\sqrt{n})$, but not necessarily to
 be optimally accurate, which leaves us some choice. For computational
 efficiency, we would require $\theta_0$ to be computationally fast and
 that it does not require an initialization itself. For global robustness
 of $S_n$, we choose $\theta_0$ to have highest possible breakdown point.

In the GPD case, little is known about the highest attainable breakdown
point. According to \cite{R:H:10a}, promising candidates for $\theta_0$
are given by so-called LD-estimators (\cite{Ma:99}),
which obtain their estimates for shape and scale by matching
empirical \textit{l}ocation and \textit{d}ispersion measures
against their respective model counterparts.
For this paper, we confine ourselves to the use of the particular
LD-estimator MedkMAD which in \cite{R:H:10a} has proven best
among all considered candidates according to both computational
efficiency and breakdown point. It is defined as follows:
as location measure we use the median, whereas as dispersion,
we use kMad, an asymmetric version of the MAD, defined
as $${\rm kMad} = \inf\{s>0:F(m+k s)- F(m-s) \geq \frac{1}{2}\}$$
Parameter $k$ reflects the skewness of the distribution, and has to
be tuned---for our purposes it suffices to take $k=10$.

\paragraph{Unknown radius}
As it is visible in \eqref{OMSEdef}, OMSE requires the radius $r$ of the
neighborhood to be known, which is almost never the case in practice.
To this end, we apply a concept from \cite{Ried:07}:
for any (arbitrarily fixed) radius $s$ and fixed procedure ${\rm OMSE}_s$
(optimal for radius $s$), we vary the true radius $r$ and determine the maximal
efficiency loss in terms of relative maxMSE in relation to the best procedure
knowing the true radius $r$ (i.e., ${\rm OMSE}_r$) and then,
in an outer loop minimize this maximal efficiency loss, varying $s$.
This gives a least favorable radius $s=r_{lf}$ for the neighborhood.
The estimator optimal on the neighborhood of this radius $r_{lf}$, i.e.,
${\rm OMSE}_{r_{lf}}$, is called  \textit{radius-minimax estimator} ({RMXE})
and is recommended.

\subsection{Software Implementation}  \label{Software}
As general software environment, we use the open source
software {\sf R}, see \cite{Rman:09}.

The solution of the implicit equations \eqref{OMSEdef} and \eqref{MBREdef}
involves numerical solution of fixed point equations as well as numerical
integration to evaluate the expectations.
A general object-oriented framework for the implementation
of these solutions can be found in {\sf R} package \texttt{ROptEst}
(\cite{K:R:09}). This package also covers RMXE.

The implementation of kMad can be found in {\sf R} package
\texttt{distrEx} (\cite{K:R:11}).
Similarly, the MedkMAD estimator has been implemented
in {\sf R} by the second author; code is available upon request.

In the GPD case, we encounter certain difficulties caused
by the lack of (complete) equivariance. For computational efficiency, the respective
Lagrange multipliers arising in MBRE, OMSE, and RMXE, therefore have been
archived for a sufficiently dense grid of $\xi$-values, so that for arbitrary
starting values of the shape coordinate of MedkMAD, the respective
Lagrange multipliers needed to compute the one-step estimator can
easily be obtained by interpolation. {\sf R}-code is again available upon request.

\section{Data Set and Evaluation of the Optimally-Robust Procedures}
\label{Sec:DataSet}

Of course, we are interested in applying these procedures to real data.
For this purpose, we use the \texttt{Algo OpData} database from
Algorithmics~Inc. \texttt{Algo OpData} contains operational losses extracted
from public data sources such as news media and the regulatory bodies.
As of July 2010, the database includes more than $12,000$ publicly reported
operational risk losses from all industry sectors. These data have been
collected in 1972--2010, majority of losses recorded within last $20$ years.
In particular, it provides detailed information about operational loss events
over one million USD from $2431$ financial institutions in compliance with
Basel~II business line and event type definition.
We use for calculations only data from the financial sector, which comprise
$5462$ losses over mostly $20$ years, not adjusted for inflation.
For practical application, the data should be scaled by an appropriate
scaling method \cite[\S 254]{Basel:10} and adjusted for inflation
\cite[\S 191]{Basel:10}, but in this paper we use the data without scaling
and inflation adjustment for illustration purposes.

Since the data is collected from public sources, due to the
thresholding/censoring mentioned in Subsection~\ref{datasit},
the severity of losses is likely to be extremal (heavy-tailed).
This makes the \texttt{Algo OpData}
different from other external operational loss data stemming from, e.g.,
the ORX database. In that sense, it is appropriate to consider the losses
unexpected---they can be used for scenario analyzes or to model the extreme
tails of severity distributions.

As required in \cite{Basel:06}, \texttt{Algo OpData} is structured as a
matrix with nine\footnote{Column `Others' contains loss data from business lines other than the ones 
defined in \cite{Basel:06}.} columns with respective business lines (BL) of the institutions
and seven rows representing the operational risk event types (ET) (see Table~1).
Here, $N$ is the total number of losses from $I$ financial institutions
over $T$ years, $n_{i,j}$ denotes the number of losses
for the ($\mathop{\rm{ET}_i,\rm{BL}_j}$) cell,
and $\hat\lambda_{i,j}$ is its average per year for a single institution, so that the following holds:
$$N = \sum_{i} \sum_{j} n_{i,j}\;, \quad n_{i,\bullet} = \sum_{j} n_{i,j}\;,
\quad  n_{\bullet,j} = \sum_{i} n_{i,j}\;,
\quad \hat\lambda_{i,j} = \frac{n_{i,j}}{I T}\;,
\quad \hat\lambda_{i,\bullet} = \frac{n_{i,\bullet}}{I T}\;,
\quad \hat\lambda_{\bullet,j} = \frac{n_{\bullet,j}}{I T}.$$

\begin{table}[t]
\centering
\begin{small}
\fbox{
\begin{tabular}{|c|c|c|c|c|c|c|c|c|c|c|c|c|c|}
\hline
&&\multicolumn{9}{|c|}{\textbf{BL}}&&&
\\
\hline
&&AS&AM&CB&CF&PS&RB&Rbrok&TS&Others&$n_{j,\bullet}$&$ \frac{n_{j,\bullet}}{N}, \ \%$&$\hat\lambda_{j,\bullet}$\\
\hline
\multirow{7}{*}{\textbf{ET}} &BDSF&&&&&10&9&4&4&9&36&1\%&0.001\\
&CPBP&51&260&171&172&46&343&329&273&570&2215&41\%&0.046\\
&DPA&&&5&&1&4&&1&15&26&0\%&0,001\\
&EPWS&1&11&20&5&&39&53&23&61&213&4\%&0.004\\
&EDPM&4&20&45&18&14&94&46&46&149&436&8\%&0.009\\
&EF&14&48&287&30&31&333&25&18&54&840&15\%&0.017\\
&IF&16&261&265&43&45&517&176&165&208&1696&31\%&0.035\\
\hline
&$n_{\bullet,i}$&86&600&793&268&147&1339&633&530&1066&\textbf{5462}&&\\
\hline
&$\frac{n_{\bullet,i}}{N}, \ \%$&2\%&11\%&15\%&5\%&3\%&25\%&12\%&10\%&20\%&&&\\
\hline
&$\hat\lambda_{\bullet,i}$&0.002&0.012&0.016&0.006&0.003&0.028&0.013&0.011&0.022&&&\\
\hline
\end{tabular}
}
\medskip
\begin{footnotesize}
\begin{tabular}{lll}
\multicolumn{2}{l}{\underline{\textit{rows:}}}\\
BDSF&Business Disruption and System Failures\\
CPBP&Clients Products and Business Practices\\
DPA&Damage to Physical Assets\\
EPWS&Employment Practices and Workplace Safety\\
EDPM&Execution Delivery and Process Management\\
EF&External Fraud\\
IF&Internal Fraud\\ \\
\end{tabular}
\begin{tabular}{lll}
\multicolumn{2}{l}{\underline{\textit{columns:}}}\\
AS&Agency Services\\
AM&Asset Management\\
CB&Commercial Banking\\
CF&Corporate Finance\\
PS&Payment and Settlement\\
RB&Retail Banking\\
Rbrok&Retail Brokerage\\
TS&Trading \& Sales 
\end{tabular}
\end{footnotesize}
\end{small}
\label{Tab:Algo_Data}
\caption{\footnotesize \texttt{Algo OpData}---the operational risk data structured
by business lines and events types according to the Basel~II requirements.}
\end{table}
\begin{floatingtable}[!trb]{
\fbox{
\begin{footnotesize}
\begin{tabular}{|l|c|c|c|}
\hline	
	\multicolumn{4}{|c|}{$\hat\lambda = 0.012$}\\
\hline	
\hline	
	Estimator&$\beta/15$&$\xi$&$\OpVaR/15$\\
\hline
\hline
	MedkMAD& $0.98$ & $1.47$ & $25.36$ \\
	MLE&	 $1.04$ & $1.28$ & $18.79$ \\
	RMXE&	 $1.01$ & $1.43$ & $24.11$ \\
	MBRE&	 $0.98$ & $1.52$ & $27.74$ \\
\hline
\end{tabular}
\end{footnotesize}
}}
\caption{\footnotesize Estimates of scale, shape of GPD,
and $1\mbox{\rm-year-}\OpVaR_{99.9\%}$ in millions USD at average number of losses per year $\hat\lambda$ for AM BL.}
\label{Tab:AM_Est}
\end{floatingtable}

For brevity we demonstrate the estimation for one BL only, i.e., Asset Management.
Taking the threshold of $u =1.6$ million USD (which gives $500$ tail events)
and applying the MedkMAD estimator (with $k=10$) to datasets from \texttt{Algo OpData},
we get starting estimates for scale and shape.
Performing a correction step with RMXE we get the final values for these parameters.
For comparison, we calculate the maximum likelihood estimator (MLE) and the MBRE.
The results of the estimation are presented in Table~2.

As indicated in Subsection~\ref{Software}, the implementation of influence functions
of RMXE and MBRE  is taken from {\sf R} package \texttt{ROptEst} and enhanced by
code of the second author, who also provides the code for MedkMAD,
while MLE is taken from {\sf R} package
\texttt{POT} (\cite{Rib:09}).

The VaR calculated with MLE is the smallest, the one calculated by MBRE is the largest.
Since the actual quantile is unknown, we cannot judge their quality without looking
the diagnostic plots given in Section~\ref{Sec:Diagnostics}.

From both theory and simulational results of  \cite{R:H:10}, it follows though that
in ideal situations, MLE is optimal, whereas in the presence of only minor
contamination MLE becomes unreliable, in which situation then OMSE and RMXE
clearly are the best choices.

This means, adding to the data single, extremely large or small losses, would
change the OpVaR value, obtained by MLE considerably, even if this added loss
is of no relevance,  whereas the value obtained through RMXE and MBRE
would only slightly change.
On the other side, in general, we have no means to decide for sure
whether a certain extreme loss is an outlier, so this loss should have influence
on the calculation of risk.
As mentioned, our optimal estimators have this property:
every observation counts, i.e.,  each observation does exert a certain,
albeit bounded influence on the estimation.

\section{Diagnostic Plots}
\label{Sec:Diagnostics}

Diagnostic plots in robust statistics aim at analyzing data for possible outliers
and their influence on the underlying estimator. We have looked at the
following diagnostic plots: influence function plots, outlyingness plots,
and QQ plots with robust confidence bands.
They should help practitioners to better understand the robust methods
when applying them to real data.

All these diagnostics are available in the {\sf R} package \texttt{ROptEst}.

\subsection{Influence Function Plots}
\label{Sec:IC}
\begin{figure}[tb]
\centering
\subfigure[Influence function of MLE]{\label{fig:MLE_IC}
\fbox{\includegraphics[scale=0.35]{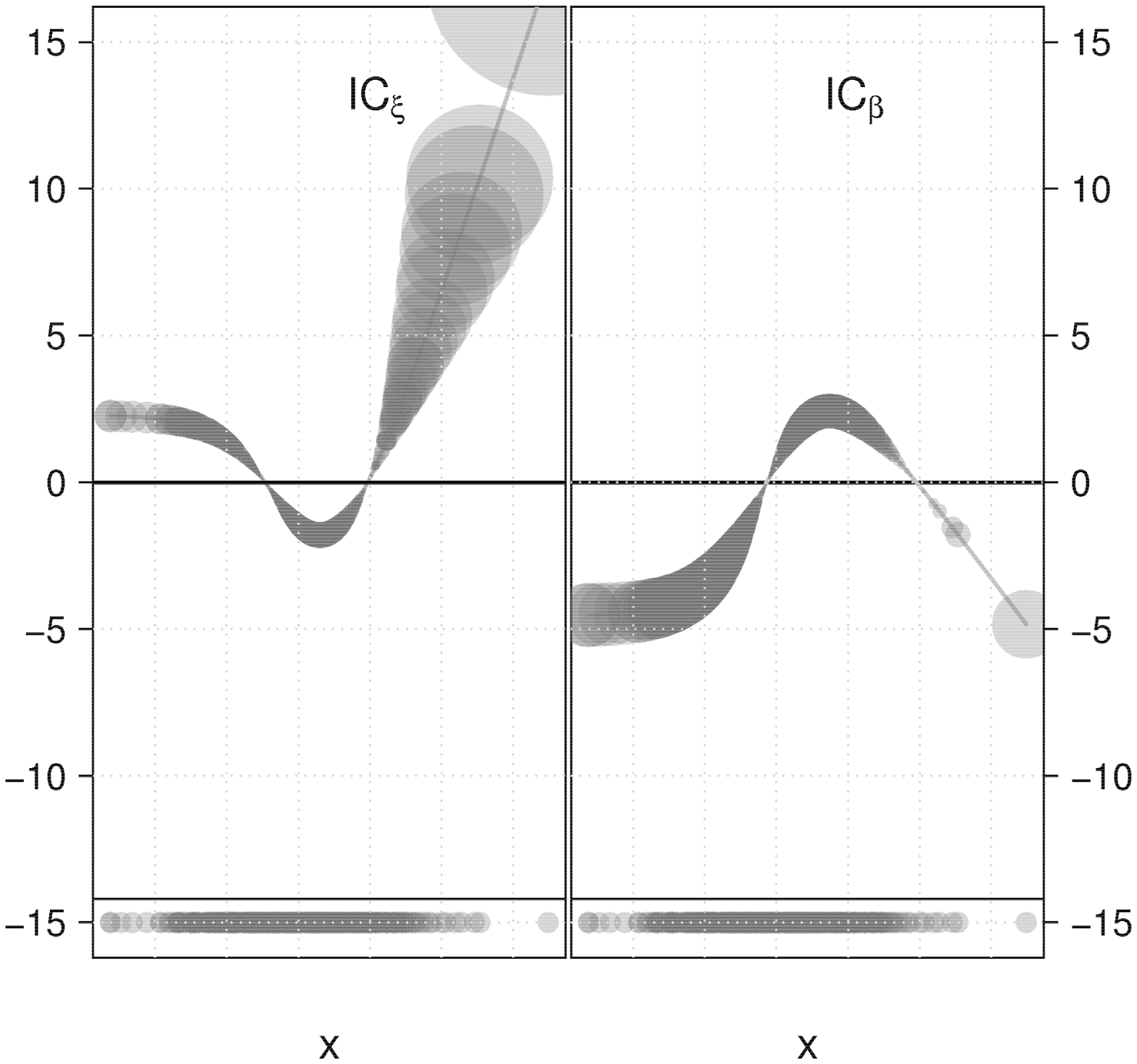}}}
\subfigure[Influence function of RMXE]{\label{fig:RMX_IC}
\fbox{\includegraphics[scale=0.35]{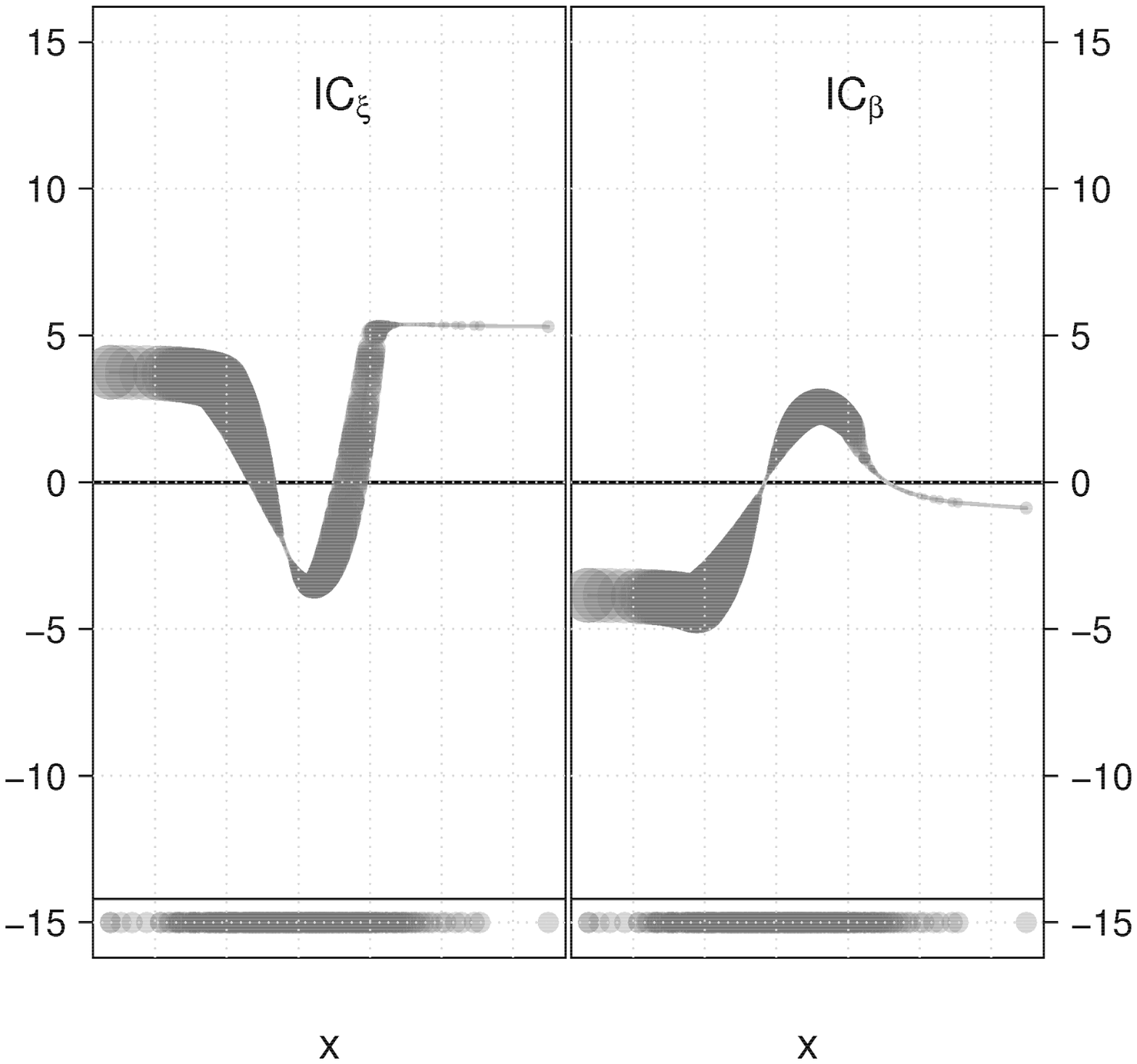}}}
\caption{\footnotesize Maximum likelihood and radius-minimax influence functions.
On the $x$-axis the values of the observations are plotted, on the $y$-axis
the respective value of the influence functions for scale and shape parameter.
The influence function for the scale parameter, $\mathop{\rm IC}_\beta$, is scaled to $\beta$ equal to one.}
\end{figure}
The influence function quantifies the (infinitesimal) influence
of each data point on the estimator. If the influence function of an
estimator is unbounded, so is the GES (see Figure~\ref{fig:MLE_IC}),
and single outliers can cause the respective estimator to
produce heavily biased estimates.
Robust estimators have bounded influence functions (e.g., RMXE in
Figure~\ref{fig:RMX_IC}).

As we estimate jointly shape and scale of GPD, the influence function has
two coordinates called influence curves, i.e.,
$\mathop{\rm IC}=(\mathop{\rm IC}_\xi, \mathop{\rm IC}_\beta)$.
On top of the lines representing  the curves themselves, we have plotted the
actual observations marked as
filled circles. The saturation of the points at the bottom of the graph
reflects the concentration of the observations, and the radius of the
points represents the size of their (joint) influence on $\xi,\beta$
in terms of $|\mathop{\rm IC}|$.

A positive [negative] value of a coordinate of the
influence function at a certain observation indicates that, infinitesimally, this
observation has increased [decreased] the respective value of
the respective parameter coordinate. Sometimes this helps in identifying
the observation(s) which has/have caused a high or low value of
the parameter estimate. Also a disequilibrium of positive and negative
values in a coordinate would be boldly visible. Without loss of generality,
assume we have much more observations with positive value in one coordinate of
the influence function, then, as the influence function must be centered,
this can only happen, if there are at least some observations with a
considerably negative influence.

As visible in the graphs, RMXE smoothly distributes the influence
of the observations, with no outstandingly influential observations
(due to boundedness). In contrast, by design, MLE cannot take
into account outliers, so considers large
observations as highly informative for parameter $\xi$,
thereby attributing high influence to some few observations at the
very right of the plot of $\mathop{\rm IC}_\xi$.
\subsection{Outlyingness Plot}
\label{Sec:OutlyingPlot}
Outlyingness plots help to detect outliers, i.e., observations which deviate
in some extent from the majority of data.

The plots discussed here translate ideas discussed in \cite{Hub:05} to our GPD case;
this case is not covered by the cited reference, as the model does neither fall
into the scope of (multivariate) location-scale type models nor is it a regression model.
Still, we follow the authors in the following two-step procedure:

In a first step, model parameters and covariances are estimated from the data
by \textit{robust techniques}. In the presence of outliers, classical estimators are
prone to \textit{masking} effects: some few large outliers
may distort our quantification of outlyingness such that other (smaller) outliers
no longer are identifiable; similarly, but less harmful in most cases, some
``clean data'' may look like outliers in the (distorted) perspective of the
outlyingness measure, an effect called \textit{swamping}. Robust procedures
 avoid both effects to large extent.

In a second step, for outlier detection, we apply an \textit{unbounded criterion}
to the data, e.g.\ the quadratic form defining the Mahalanobis norm.
This unboundedness helps to discern outliers properly, which in a bounded
criterion would become indistinguishable from non-outliers.
However, where model parameters and covariances are needed to evaluate
this criterion, e.g.\ the covariance to determine the Mahalanobis norm,
we use the robust ones from the first step.

Usually to visualize outlyingness, two criteria from the second step
are used in parallel---one of for the $x$- one for the $y$-axis.
In each coordinate a threshold (preferably a suitable high quantile)
is chosen, giving a partition into four quadrants. Observations
 simultaneously falling beyond both thresholds are flagged as outliers,
which, of course, must be seen as only an indication for being an outlier,
as both usual error-types of a test may occur.

\begin{figure}[!tb]
\centering
\subfigure[Outlyingness Plot]{\label{fig:outlyingPlot}
\fbox{\includegraphics[scale=0.3,angle=270]{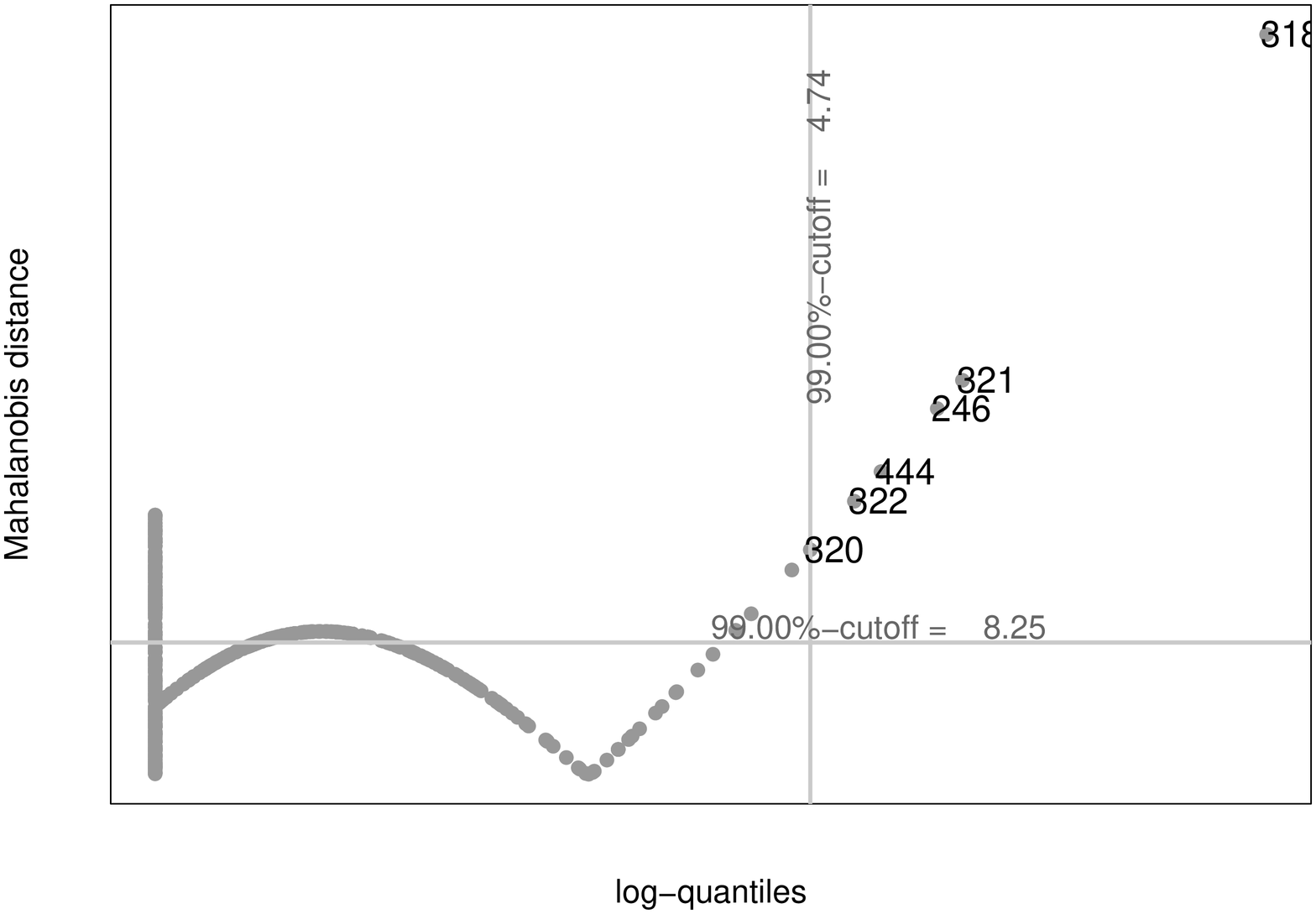}}}
\subfigure[QQ Plot]{\label{fig:qqPlot}
\fbox{\includegraphics[scale=0.3,angle=270]{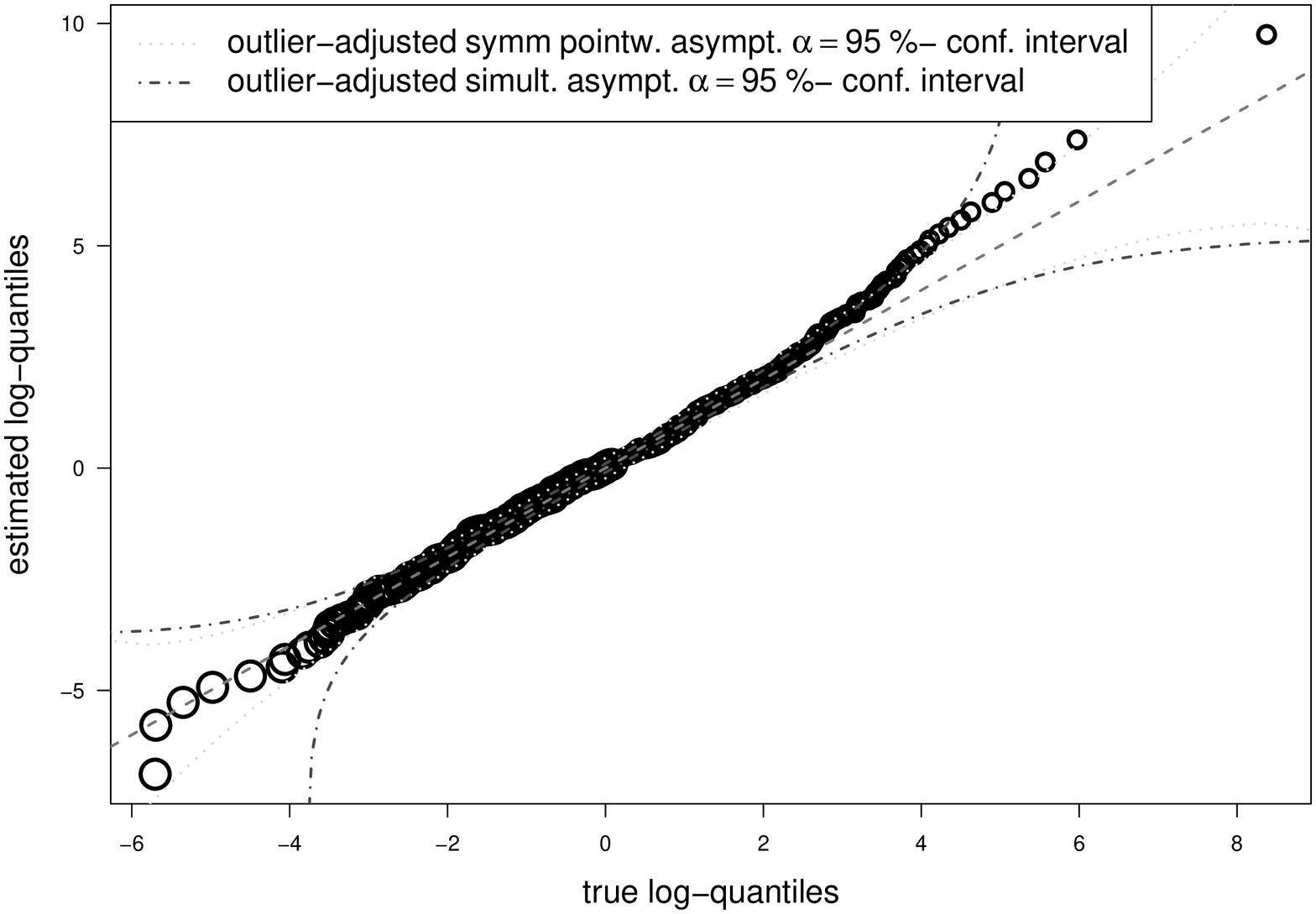}}}
\caption{\footnotesize Diagnostic plots: outlyingness plot and QQ plot with robust confidence bands}
\end{figure}

There are different variations of outlying plots:
 distance-distance, distance-projection, and projection-projection plots.
Our outlyingness plot for the GPD is a distance-projection plot,
 which for parameter estimation uses RMXE and for covariances the Minimum
 Covariance Determinant (MCD) estimator from \cite{Rou:LMS},
as implemented in {\sf R} package \texttt{rrcov}, see \cite{Tod:09}.
More precisely, we plot a (robustified, empirical) Mahalanobis distance
  of the MLE influence function against the usual data quantiles.
This gives Figure~\ref{fig:outlyingPlot}.
We use thresholds given by the $99\%$ quantile of the
$\chi_2^2$-distribution with non-centrality $0$
on the $y$-axis and $99\%$ quantile of the data on the $x$-axis.

Table~\ref{Tab:Outlier_Events} shows which operational losses in
the Asset Management BL are flagged as outliers in Figure~\ref{fig:outlyingPlot}.
Of the six outlying losses, indexed as $246$, $318$, $320$, $321$, $322$, $444$,
four are caused by the recent fraud by B.~L. Madoff and the remaining one resulted from
a Ponzi scheme fraud.
Although these losses are probably outliers, they should be included
into the estimation instead of being skipped, as they could also carry
some valuable information for future losses.
Classical MLE however interprets these values as ``usual observations''
and, as a consequence, assigns them too much influence, no matter whether
their relevance or reproducibility is doubtful or not.
Robust RMXE, includes these doubtful observations too, but downweighs them,
so that their influence on the resulting estimates is smaller than those of
the remaining losses (see Table~\ref{Tab:Weights}).

\subsection{QQ Plot With Robust Confidence Bands}

Quantile-quantile (QQ) plots aim at visualizing the quality of a model fit:
empirical quantiles of the observations are plotted against
the quantiles of the fitted model distribution. A concentration
of the plotted points around the line $y=x$ indicates a high quality,
while large deviations indicate outliers or a failure of the model
fit.

Still, there is estimation uncertainty in the data, which
can be captured by suitable confidence intervals grouped to bands
according to their position, larger [narrower] bands indicating
higher [lower] uncertainty.

As usual in this context, there are both pointwise and simultaneous
confidence bands. Pointwise confidence intervals describe the stochastic variability of
the empirical distributions of the data for each quantile individually,
while simultaneous confidence bands capture the variability of the whole
empirical cumulative distribution function (ecdf), so that, on average,
$95\%$ of the graphs produced by ecdfs will completely lie within these bounds.

Taking outlier-induced model deviations into account,
for robust confidence bands the nominal confidence level has to be adjusted
accordingly: to warrant a nominal level $\alpha$ we have to increase
the defining level to $\alpha+r/\sqrt{n}$.

The QQ plot of RMXE-estimated GPD quantiles versus real quantiles is depicted
in Figure~\ref{fig:qqPlot}. The size of the points reflects their weight
in the influence function, so that downweighed observations get smaller
circles.  One can see that the fit is good in the lower and middle quantiles
where (at least in the middle) also model uncertainty is low, but poorer in
the upper ones around $4$, where the points even fall outside the (simultaneous)
confidence bands.
This phenomenon appears to be due to the outlying data points in the tails
(that at least get downweighed by RMXE). The widening of the confidence bands at the lower and upper
ends is common and caused by the little empirical evidence available in this
area.

\begin{table}[!htb]
\centering
\fbox{
\begin{footnotesize}
\begin{tabular}{|c|c|c||c|c|c||c|c|c|}\hline
\multicolumn{1}{|c}{\begin{tabular}{l}Obs. \\ Index\end{tabular} }
&\multicolumn{1}{|c}{\begin{tabular}{l}Loss Value \\(billions\footnotemark USD )  \\  \end{tabular}}
&\multicolumn{1}{|c||}{Weight}
&\multicolumn{1}{ c}{\begin{tabular}{l}Obs. \\ Index\end{tabular} }
&\multicolumn{1}{|c}{\begin{tabular}{l}Loss Value \\(billions USD) \\   \end{tabular}}
&\multicolumn{1}{|c||}{Weight}
&\multicolumn{1}{c}{\begin{tabular}{l}Obs. \\ Index\end{tabular} }
&\multicolumn{1}{|c}{\begin{tabular}{l}Loss Value \\(billions USD) \\   \end{tabular}}
&\multicolumn{1}{|c|}{Weight}\tabularnewline
\hline
\hline

$246$&$ 6.0 $&$0.18$ &  $320$&$ 2.4$&$0.24$& $322$  &$ 3.3$&$0.21$\tabularnewline
$318$&$ 65.0 $&$0.11$&  $321$&$ 7.2$&$0.17$ &$444$  &$ 4.0$&$0.20$\tabularnewline
\hline
\end{tabular}\end{footnotesize}
}
\caption{\footnotesize
Weights of outliers in RMXE with corresponding loss values
}

\label{Tab:Weights}
\end{table}
\footnotetext{$1$ billion = $10^9$}
\begin{table}[!tbrl]
\begin{scriptsize}
\fbox{
\begin{tabular}{|c|c|c|c|c|c|c|}
\hline
\multirow{4}{*}{\begin{tabular}{l}Outlier \\\\ Index \end{tabular}}
&\multirow{4}{*}{\begin{tabular}{l}\\\\ Business Line\end{tabular}}
&\multirow{4}{*}{\begin{tabular}{l}\\\\ Event Type\end{tabular}}
&\multirow{4}{*}{\begin{tabular}{l}Organization\end{tabular}}
&\multirow{4}{*}{\begin{tabular}{l}Loss Amount \\\\ (billions USD)\end{tabular}}
&\multirow{4}{*}{\begin{tabular}{l}Settlement \\\\ Date\end{tabular}}
&\multirow{4}{*}{Location} \\
&&&&&&\\
&&&&&&\\
&&&&&&\\
\hline
\hline
246 & Asset Management & \begin{tabular}{l}Clients Products \\ and Business Practices\end{tabular}
  & Amaranth Advisors & 6.0 & 9/18/2006 & \begin{tabular}{l} North America \\ Canada \\ Alberta\end{tabular}
\\
\hline
318 & Asset Management & Internal Fraud  &
\begin{tabular}{l} Bernard Madoff Investment \\ Services LLC \end{tabular}
& 65.0 & 12/11/2008 & \begin{tabular}{l} North America \\ United States \\ New
York \end{tabular} \\
\hline
320 & Asset Management & External Fraud  & Ascot Partners L.P. &
2.4 & 12/16/2008 & \begin{tabular}{l} North America \\ United States \\ New York\end{tabular} \\
\hline
321 & Asset Management & External Fraud  & Fairfield Greenwich Group &
7.2 & 12/15/2008 & \begin{tabular}{l} North America \\ United States \\ Connecticut\end{tabular} \\
\hline
322 & Asset Management & External Fraud  & MassMutual Financial Group &
3.3 & 12/16/2008 &  \begin{tabular}{l}North America \\ United States \\ New York\end{tabular} \\
\hline
444 & Asset Management & Internal Fraud  & Cash Plus & 4.0 &
10/9/2009 &  \begin{tabular}{l}\hspace*{-4ex}Caribbean \\ \hspace*{-4ex}Jamaica\end{tabular} \\
\hline
\end{tabular}
}
\end{scriptsize}
\caption{\footnotesize Outlying events in Asset Management business line}
\label{Tab:Outlier_Events}
\end{table}

\section*{Conclusion}

This article applies optimally-robust estimation techniques to
real world data for the calculation of the regulatory capital
for operational risks within the LDA (AMA) setting, according
to Basel~II requirements.
The data we use is taken from the \texttt{Algo OpData} database of
Algorithmics~Inc. No scaling has been
applied, so the results we obtain are only meant for illustrative
purposes.

Still, all other steps required in LDA have been gone through:
we model the severity of tail events by a GPD distribution and the
frequency of losses with a Poisson distribution, and apply a
single-loss approximation for the corresponding $99.9\%$ quantile
of the compound loss distribution.
For estimation of the GPD parameters, we focus on respective
optimally-robust estimators, OMSE, OBRE, and RMXE, in their
specialization to the GPD case taken from \cite{R:H:10}
where they are also compared with several competitors but
as predicted by theory turn out optimal even at sample sizes
down to $40$.
For these estimators, we use a robust starting estimator, MedkMAD, based
on the median and the asymmetric median of absolute deviations.
Its qualification as globally robust, computationally efficient
starting estimator has been taken from \cite{R:H:10a}.

In evaluating our estimators we have found no difficulties.
In case of business line Asset Management, our robust estimators
indicate the need of a higher regulatory capital
than indicated by classical MLE ($28\%$ higher for RMXE),
and a value of $28\%$ for the relative deviation indicates the
presence of influential outliers.
A statement of the type ``robustly estimated OpVaR is generally higher than the one
obtained by classical methods'' however is not true. The order
varies from business line to business line.

To assess the quality of our robust estimates and the respective
model fit at real data, and to discern potential outliers,
we present robust diagnostic plots.
At the present data set, our outlyingness plot was able to grasp
the singular pattern of the Madoff fraud. For the majority of the
data, however, the robust model fit according to the QQ plot
seems reasonably good. In the influence function plot, we see
that at the actual data, in particular the shape parameter is concerned
with highly influential observations in the MLE case, whereas
no such pattern is visible in the RMXE case.

For the evaluation of the respective estimators, as well as for
the diagnostic plots, we use publicly available software provided in the {\sf R}
package \texttt{ROptEst}, tuned for computational efficiency
with own code, as well as own routines for the computation of MedkMAD;
the code is available upon request.

\section*{Acknowledgement} We thank two anonymous referees for their helpful
and valuable comments. All three authors have equally contributed to the present
paper.

\def\textSc#1{#1}
\begin{footnotesize}

\end{footnotesize}


\begin{thebibliography}{00}
\setlength{\itemindent}{-2mm}
\setlength{\itemsep}{.4mm plus1.mm minus.2mm}

\bibitem[Balkema and de Haan(1974)]{B:H:74}
\textSc{Balkema, A. and de Haan, L.} Residual life time at great age. \textit{Annals of Probability} {2}, 792--804, 1974.

\bibitem[Basel~II(2006)]{Basel:06}
\textSc{Basel Committee on Banking Supervision.} International Convergence of Capital Measurement and Capital Standards: A Revised Framework. \url{http://www.bis.org/publ/bcbs128.pdf}, June, 2006.

\bibitem[BIS(2010)]{Basel:10}
\textSc{Basel Committee on Banking Supervision.}
Operational Risk -- Supervisory Guidelines for the Advanced Measurement Approaches. \textit{Consultative Document}.  \url{http://www.bis.org/publ/bcbs184.pdf}, December, 2010.

\bibitem[BIS(2010a)]{Basel:10a}
\textSc{Basel Committee on Banking Supervision.}
2010 FSI Survey on the Implementation of the New Capital Adequacy Framework: Summary of responses to the Basel~II implementation survey. Occasional Paper, N$^o$9.  \url{http://www.bis.org/fsi/fsipapers09.pdf}, August, 2010

\bibitem[Beirlant et al.(1999)]{B:D:G:M:99}
\textSc{Beirlant, J., Dierckx, G., Goegebeur, Y., and Matthys, G.} Tail index estimation and an exponential regression model. \textit{Extremes} {2}, 177--200, 1999.

\bibitem[Beirlant et al.(1996)]{B:V:T:96}
\textSc{Beirlant, J., Vynckier, P., and Teugels, J.~L.} Tail index estimation, Pareto quantile plots, and regression diagnostics. \textit{J. Amer. Statist. Assoc.} {91}, 1659--1667, 1996.

\bibitem[B\"ocker and Kl\"uppelberg(2005)]{B:K:05}
\textSc{B\"ocker, K. and Kl\"uppelberg, C.} Operational VAR: a Closed-Form Approximation. \textit{RISK Magazine}, December, 90--93, 2005.


\bibitem[Brazauskas and Kleefeld(2009)]{B:K:09}
\textSc{Brazauskas, V. and Kleefeld, A.} Robust and Efficient Fitting of the Generalized Pareto Distribution with Actuarial Applications in View. \textit{Insurance: Mathematics and Economics} {45}, 424--435, 2009.

\bibitem[Chavez-Demoulin et al.(2006)]{Chav:06}
\textSc{Chavez-Demoulin, V., Embrechts, P. and Neslehova, J.} Quantitative models for operational risk: extremes, dependence and aggregation. \textit{Journal of Banking and Finance} {30}(10), 2635--2658, 2006.

\bibitem[Chernobai and Rachev(2006)]{C:R:06}
\textSc{Chernobai, A. and Rachev, S.~T.} Applying Robust Methods to Operational Risk Modelling. \textit{Journal of Operational Risk} {1}(1), 27--41, 2006.

\bibitem[Chernobai et al.(2011)]{C:J:Y:11}
\textSc{Chernobai, A., Jorion, P. and Yu, F.} The Determinants of Operational Risk in U.S. Financial Institutions. \textit{Journal of Financial and Quantitative Analysis}. Forthcoming.

\bibitem[Cope and Labbi(2008)]{C:L:08}
\textSc{Cope, E. and Labbi, A.} Operational Loss Scaling by Exposure Indicators: Evidence from the ORX Database. \textit{Journal of Operational Risk} 3(4), 2008.

\bibitem[De Fontnouvelle et al.(2006)]{deFon:06}
\textSc{De Fontnouvelle, P., De Jesus-Rueff, V., Jordan, J. and Rosengren, E.} Capital and Risk: New Evidence on Implications of Large Operational Losses. \textit{Journal of Money, Credit, and Banking} {38}(7), 2006.

\bibitem[De Fontnouvelle et al.(2007)]{deFon:07}
\textSc{De Fontnouvelle, P., Rosengren, E. and Jordan, J.}
Implications of Alternative Operational Risk Modelling Techniques,
In \textit{Carey, M. and Stulz, R.~M. The Risks of Financial Institutions},
University of Chicago Press, 475--512, \url{http://www.nber.org/books/care06-1}, 2007.

\bibitem[Dell'Aquila and Embrechts(2009)]{D:E:09}
\textSc{Dell'Aquila, R. and Embrechts, P.} Extremes and robustness: a contradiction? \textit{Financial Markets and Portfolio Management} {20}, 103--118, 2006.

\bibitem[Dupuis(1998)]{Du:98}
\textSc{Dupuis, D.~J.}
{Exceedances over high thresholds: A guide to threshold selection.}
\textit{Extremes} {1}(3), 251--261, 1998.

\bibitem[Dupuis and Field(1998)]{D:F:98}
\textSc{Dupuis, D.~J. and Field, C.~A.} Robust estimation of extremes. \textit{Canad. J. Statist.} {26}(2), 199--215, 1998.

\bibitem[{Dupuis and Morgenthaler}(2002)]{D:M:02}
\textSc{Dupuis, D.~J. and Morgenthaler S.}
{Robust weighted likelihood estimators with an application to bivariate extreme value problems.}
\textit{Canad. J. Statist.} {30}(1), 17--36, 2002.

\bibitem[Dupuis and Victoria-Feser(2006)]{D:V-F:06}
\textSc{Dupuis, D.~J. and Victoria-Feser, M.-P.}
A robust prediction error criterion
for Pareto modelling of upper tails.
\textit{Canad. J. Statist.} {34}(4), 639--658, 2006.

\bibitem[Embrechts et al.(2003)]{Emb:03}
\textSc{Embrechts, P., Furrer, H. and Kaufmann, R.}
Quantifying Regulatory Capital for Operational Risk Derivatives Use. \textit{Trading \& Regulation} {9}(3), 217--233, 2003.

\bibitem[Fernholz(1983)]{Fe:83}
\textSc{Fernholz, L.T.},
{Von Mises Calculus for Statistical Functionals}. \textit{Lecture Notes in Statistics},
vol.~19, Springer, New York, 1983.

\bibitem[Field and Smith(1994)]{F:S:94}
\textSc{Field, C. and Smith, B.}
{Robust Estimation---A Weighted Maximum Likelihood Estimation.}
{International Review} {62}(3), 405--424, 1994.

\bibitem[{Hampel et al.}(1986)]{Ha:Ro..:86}
\textSc{Hampel, F.~R., Ronchetti, E.~M., Rousseeuw, P.~J. and Stahel, W.~A.}
\textit{Robust statistics. The approach based on influence functions.}
{Wiley,}  1986.

%

\bibitem[Huber(1964)]{Hu:64}
\textSc{Huber, P.~J.}
{Robust estimation of a location parameter}.
\textit{Ann. Math. Statist.}{35}, 73--101, 1964.

\bibitem[Huber(1981)]{Hu:81}
\textSc{Huber, P.~J.}
\textit{Robust statistics},
{Wiley}, 1981.

\bibitem[Hubert et al.(2005)]{Hub:05}
\textSc{Hubert, M., Rousseeuw, P. J. and Van Aelst, S.} Multivariate Outlier Detection and Robustness.
\textit{Handbook of Statistics, Volume 23: Data Mining and Computation in Statistics} (C.~R. Rao, E.~J. Wegman, J.~L. Solka, Eds.), Elsevier, pp. 263--302, 2005.

\bibitem[Ju\'arez and Schucany(2004)]{J:S:04}
\textSc{Ju\'arez, S.~F. and Schucany, W.~R.}
{Robust and Efficient Estimation for the Generalized Pareto Distribution.}
\textit{Extremes} {7}(3), 237--251, 2004.



\bibitem[{Kohl et al.}(2010)]{K:R:R:09}
\textSc{Kohl, M., Rieder, H., and Ruckdeschel, P.}
{Infinitesimally Robust Estimation in General Smoothly Parametrized Models.}
\textit{Stat. Methods Appl.}, {19}, 333--354, 2010.

\bibitem[Kohl and Ruckdeschel(2011)]{K:R:09}
\textSc{Kohl, M. and Ruckdeschel, P.} \texttt{ROptEst}: Optimally robust estimation. {\sf R} package available in version 0.8 on CRAN, \url{http://cran.r-project.org}, 2011.

\bibitem[Kohl and Ruckdeschel(2011)]{K:R:11}
\textSc{Kohl, M. and Ruckdeschel, P.}
\texttt{distrEx}: Extensions of package distr and some additional functionality.
{\sf R} package available in version 2.3 on CRAN, \url{http://cran.r-project.org}, 2011.

\bibitem[{Marazzi and Ruffieux}(1999)]{Ma:99}
\textSc{Marazzi, A. and Ruffieux, C.}
{The truncated mean of asymmetric distribution.}
\textit{Computational Statistics \& Data Analysis}, {32}, 79--100, 1999.

\bibitem[{Maronna et al.}(2006)]{M:M:Y:06}
\textSc{Maronna, R.~A., Martin, R.~D. and Yohai, V.~J.}
\textit{Robust Statistics: Theory and Methods.}
Wiley, 2006.

\bibitem[Marshall(2001)]{Marsh:01}
\textSc{Marshall, L.~C.} \textit{Measuring and Managing Operational Risks in Financial Institutions. Tools, Techniques, and other Resources}. Wiley, 2001.


\bibitem[Moscadelli(2004)]{Mosc:04}
\textSc{Moscadelli, M.} The modelling of operational risk: experience with the analysis of the data collected by the Basel committee. Technical Report 517, Banca d'Italia, 2004.

\bibitem[Neslehova et al.(2006)]{Nesl:06}
\textSc{Neslehova, J., Embrechts, P. and Chavez-Demoulin, V.}
Infinite mean models and the LDA for operational risk.
\textit{Journal of Operational Risk} {1}(1), 3--25, 2006.


\bibitem[Peng and Welsh(2001)]{P:W:01}
\textSc{Peng, L. and Welsh, A.~H.}
{Robust Estimation of the Generalized Pareto Distribution. }
\textit{Extremes} {4}(1), 53--65, 2001.

\bibitem[Pickands(1975)]{Pick:75}
\textSc{Pickands, J.} Statistical Inference Using Extreme Order Statistics.
\textit{Annals of Statistics} {3}(1), 119--131, 1975.

\bibitem[{{\sf R} Development Core Team}(2011)]{Rman:09}
\textSc{{\sf R} Development Core Team}. {\sf R}: A language and environment for
  statistical computing. {\sf R} Foundation for Statistical Computing,
  Vienna, Austria.
  ISBN 3-900051-07-0, \url{http://www.R-project.org}, 2011.

\bibitem[Ribatet(2009)]{Rib:09}
\textSc{Ribatet, M.} \texttt{POT}: Generalized Pareto Distribution and Peaks over Threshold.
{\sf R} package, version 1.1-0, \url{http://cran.r-project.org}, 2009.

\bibitem[Rieder(1994)]{Ried:94}
\textSc{Rieder, H.} \textit{Robust Asymptotic Statistics.} Springer, 1994.

\bibitem[Rieder et al.(2008)]{Ried:07}
\textSc{Rieder, H., Kohl, M. and Ruckdeschel, P.}
The cost of not knowing the radius.
\textit{Statistical Methods \& Applications} {17}(1), 13--40, 2008.

\bibitem[Rousseeuw(1984)]{Rou:LMS}
\textSc{Rousseeuw, P.J.}
 Least Median of Squares Regression.
\textit{J. Amer. Statist. Assoc.} {79}(388), 871--880, 1984.


\bibitem[{Ruckdeschel}(2006)]{R:06}
\textSc{Ruckdeschel, P.}
{A motivation for $1/\!\sqrt{n}$-shrinking-neighborhoods.} \textit{Metrika},
{63}(3) 295--307, 2006.

\bibitem[Ruckdeschel and Horbenko(2011)]{R:H:10a}
\textSc{Ruckdeschel, P. and Horbenko, N.} Yet another breakdown point notion:
EFSBP --illustrated at scale-shape models. ArXiv 1005.1480
, 2011.

\bibitem[Ruckdeschel and Horbenko(2010)]{R:H:10}
\textSc{Ruckdeschel, P. and Horbenko, N.} Robustness Properties of Estimators in Generalized Pareto Models. Technical Report ITWM N$^o$182, \url{http://www.itwm.fraunhofer.de/fileadmin/ITWM-Media/Zentral/Pdf/Berichte_ITWM/2010/bericht_182.pdf}, 2010.


\bibitem[Todorov(2009)]{Tod:09}
\textSc{Todorov, V.} \texttt{rrcov}: Scalable Robust Estimators with High Breakdown Point. {\sf R} package, version 1.0, \url{http://cran.r-project.org}, 2009.

\bibitem[Vandewalle et al.(2007)]{V:B:C:H:07}
\textSc{Vandewalle, B., Beirlant, J., Christmann, A., and Hubert, M.}
A robust estimator for the tail index of Pareto-type distributions.
\textit{Comput. Statist. \& Data Anal.} {51}(12), 6252--6268, 2007.



\bibitem[{van der Vaart}(1998)]{VdW:98}
\textSc{van der Vaart, A.W.}
\textit{Asymptotic statistics.}
{Cambridge Univ. Press}, {1998}.

\bibitem[Witting(1985)]{Wit:85}
\textSc{Witting, H.}
\textit{Mathematische Statistik I: Parametrische Verfahren bei festem Stichprobenumfang}.
B.G.~Teubner, Stuttgart, 1985.


\end{thebibliography}
\end{document}